\documentclass[11pt,a4,fleqn]{article}

\usepackage{authblk}
\usepackage{xcolor}
\usepackage{amsmath}
\usepackage{graphicx}
\usepackage[nosort]{cite}
\usepackage{placeins} 
\usepackage{gensymb}
\usepackage{array}
\usepackage{subfig}
\usepackage{epsfig}
\usepackage[margin=1.5in]{geometry}
\usepackage{hyperref}

\usepackage{fancyhdr}
 
\usepackage{setspace}
\title{\textbf{Unsteady Navier-Stokes studies on loads, wake and dynamic stall characteristics of a two-bladed vertical axis wind turbine}}

\author[ ]{Galih Bangga}
\author[ ]{Thorsten Lutz}
\author[ ]{Amgad Dessoky}
\author[ ]{Ewald Kr{\"a}mer}
\affil[ ]{Institute of Aerodynamics and Gas Dynamics (IAG), University of Stuttgart, Pfaffenwaldring 21 Stuttgart 70569, Germany}
\affil[ ]{\textit{bangga@iag.uni-stuttgart.de and galih.bangga90@gmail.com}}
\date{}                     
\begin{document}
\maketitle

\abstract{

\noindent Computational fluid dynamics (CFD) studies are carried out on a two-bladed vertical axis wind turbine (VAWT) operating at a wind speed of 8 m/s for the tip speed ratios ($\lambda$) of 0.50 - 3.0. The blade consists of the NACA 0021 airfoil with the chord length of 0.265 m and a rotor radius of 1 m. Basic sensitivity studies for various time step sizes are carried out. The results are validated against available measurement data from literature. An excellent agreement is obtained for small $\lambda$ up to optimum condition. For the higher tip speed ratios, the two dimensional CFD computations predict higher results than the wind tunnel experiment, but they are very similar to the field measurement data. Wake characteristics are presented in the present studies, showing that the wake becomes Gaussian at 5 times radius downsteam of the rotor. It is shown that complex flow phenomena occur due to dynamic stall onset especially for the smaller tip speed ratio.}

\FloatBarrier
\section{Introduction}
\label{intro}

The growing demand for energy and the intensification of global climate change, the social development, and human survival are under a great threat due to shortage of fossil energy \cite{song}. Alternatives for the energy sources need to be identified. Wind energy has received a great attention by renewable energy communities in the recent years due to its high potential in generating long term sustainable energies. This type of energy source has become the fastest growing industry for renewable energy with more than 30\% annual growth rate \cite{hutomo}. 

Currently, there are two categories of modern wind turbines according to the axis of rotation: (1) Horizontal Axis Wind Turbines (HAWTs) and (2) Vertical Axis Wind Turbines (VAWTs), which are continuously developed. Nowadays, there is a strong interest from the wind energy community to harvest the energy within built environments at urban, sub-urban and remote areas in addition to large wind farms. HAWTs are not economically and socially suitable for such environments as they need large space to build, and a sophisticated yaw mechanism is required because wind direction often changes significantly \cite{bangga1}. On contrary, VAWT offers a simpler design than HAWT. The generator can be mounted on the ground, reducing the weight of the turbine that leads to a longer expected life time. Furthermore, the rotor has a small dependency towards wind direction so that no yaw mechanism is required. These advantages alleviate the manufacturing and maintenance costs of the rotor.

Despite the mentioned superiorities, it is well known already that the flow field surrounding a VAWT rotor is complex involving a strong unsteadiness. During each rotor revolution, the blade sees varying local inflow velocity and angle of attack that can manifest as dynamic stall \cite{hutomo, bangga1}. This effect has a huge influence to the aerodynamic performance of the rotor. It has been shown by Bangga et al. \cite{bangga1} that the generated power of a single bladed VAWT was negative for almost 22\% of the azimuth angle range due to the dynamic stall effect. Laneville and Vitecoq \cite{laneville} emphasized that dynamic stall presents for VAWT operating at tip speed ratio ($\lambda$) less or equal than 4, which is supported by the measurement carried out by Akins et al. \cite{akins}.

Development of vertical axis wind turbines for the time being is focused on the small scale rotors. The rotor is usually proposed as an energy alternative in a region where electricity is not available like in the forest or where HAWT is hard to build like in the middle of cities. Several measurements, wind tunnel and field test, were carried out and documented in literature \cite{kjellin, li, ferreira, dabiri, akins}. Kjellin et al. \cite{kjellin} performed a field test measurement of a 12 kW straight bladed vertical axis wind turbine. The results of the experiment was intended to be used as parameters in the control system of the turbine, in order to maintain the operation at optimum tip speed ratio. Li et al. \cite{li} conducted both field and wind tunnel experiments focusing on the influences of pitch angle, Reynolds number and turbulence intensity on the power performance of a VAWT. The outcome shows various datasets of the power curve which are valuable for the validation tests. The effect of wind skew on VAWT was investigated in detail by Ferreira et al. \cite{ferreira}. It was shown that the tip vortices influence, generated in the downwind blade passage, on the vortices generated upwind is stronger than in the non-skewed case. Dabiri \cite{dabiri} suggested to design wind turbine arrays consisting of counter-rotating turbines to enhance the power production of the wind field.

Despite the advantages of the experimental measurement, the technique cannot be widely used for its weakness of high cost, long cycle and poor flexibility \cite{song}. There are a couple of low order simulation approaches for VAWT. The well known Blade Element Momentum (BEM) is one of them. This method discretizes the blade into several sections and the resulting forces are calculated based on the provided aerodynamic polars on the corresponding sections. The main drawback of the approach is that most flow physics are neglected such as the 3D effects, and thus the accuracy depends strongly upon the provided polars \cite{bangga5}. It has been documented by Li and {\c{C}}al{\i}{\c{s}}al \cite{lied1} that the momentum method \cite{camporeale} and the boundary-element method \cite{calcagno} are often used to predict power output from a turbine. Vortex methods are developed to deal with the issue of low order simulation approaches in the physical modelling by considering not only the 1D momentum theory but also the wake induced effects on the resulting rotor loads. In \cite{lied2}, Li and {\c{C}}al{\i}{\c{s}}al extended the numerical model to simulate twin-turbine system.

Due to recent advances in high performance computations, computational fluid dynamics (CFD) approaches were applied in many engineering fields including VAWT with satisfactory results \cite{bangga1, song, ferreira2, castelli}. Ferreira et al. \cite{ferreira2} presented 2D VAWT studies for various turbulence models, showing that Detached Eddy Simulations (DES) results are not only able to predict the generation and shedding of vorticity and it’s convection, but also show an acceptable sensitivity to grid refinement (both space and time) inferring that it can be used as a suitable comparison where validation data is limited or non existent. Song et al. \cite{song} studied four meshing strategies for VAWT simulations. It has been shown that the required mesh for VAWT consists of 300 grid points on the airfoil surface with 15 cells within the boundary layer. The step size of $T/240$, where $T$ denotes period of the rotor revolution, was observed to be reasonably small enough in capturing the loads. They obtained good results using 2D URANS simulations. Recently, Bangga et al. \cite{bangga1} conducted VAWT simulations using considerably finer mesh of 500 nodes on the airfoil surface in investigating the dynamic stall influence on a single bladed Darrieus rotor operating at $\lambda$ = 2.0.

The present studies focus on the 2D modelling of a two-bladed vertical axis wind turbine rotor using high fidelity CFD simulations, aiming to obtain deeper insights into the time accurate solutions of the rotor at various tip speed ratios. At first, preliminary studies into the time scale effect on the unsteady simulations will be carried out. Then, the results are validated against available measurement database from Li et al. \cite{li}. Wake profiles are evaluated to quantify the momentum mixing of the flow in the wake area. The flow field characteristics and development of the shedding vortices surrounding the rotor will be presented and discussed in details.


\section{Methodology} \label{Sect2}
\subsection{Studied Turbine and Test Case} \label{Test}
In exploring the power performance of a vertical axis wind turbine, very recently in 2016 Li et al. \cite{li} performed wind tunnel and field test measurements of a straight bladed rotor under the influences of pitch angle, Reynolds number and wind speed. The wind tunnel experiments were carried out at the open test section of a circular type wind tunnel in Mie University, Japan \cite{li}. The wind tunnel outlet diameter and length are 3.6 m and 4.5 m, respectively, with the maximum wind speed of 30 m/s. The studies were, however, carried out at the wind speed of 8.0 m/s with less than 0.5\% turbulence intensity \cite{li}. More detailed information of the wind tunnel was given in \cite{maeda, li}. In the field test measurement, the rotor was mounted at a height $h$ = 5.0 m from the ground.

The rotor consists of two blades employing the NACA0021 airfoil cross-section rotating in clockwise direction. It shall be noted that the 2D simulations represent the mid-span section of the blade where the spanwise flow is not modelled. The turbine radius is 1.0 m, blade length is 1.2 m and the chord length is 0.265 m. In this study, a constant wind speed of 8 m/s and pitch angle of 6\degree{} were applied. A complete range of the power coefficient ($C_{power}$) curve was obtained by varying the rotational speed of the rotor. In wind turbine design, it is a common rule to specify a constant wind speed for various operating tip-speed-ratio ($\lambda$) to obtain a full power curve of the rotor. Usually, the employed wind speed is the designed rated condition that will be used in the experiment or depending on the field/wind farm situation. A similar technique was employed in obtaining the reference experimental data \cite{li}. Since the measurement data from Li et al. \cite{li} are used to validate the results, thus it is important to apply the same condition as in the experiment and the present test case was chosen based on this consideration. The simulations were carried out for eight different tip speed ratios, ranging from 0.5 up to 3.0.

\subsection{Computational Mesh and Numerical Setup} \label{Setup}
Similar to preceding studies carried out by Castelli et al. \cite{castelli}, Song et al. \cite{song} and Bangga et al. \cite{bangga1}, some detailed rotor components like its main shaft, bolts, and member bars that connect blades with main shaft were omitted in the simulations. Two dimensional geometry was applied in the studies as the blade section is exactly the same along the blade length, neglecting the influence of tip vortices.

\begin{figure}[!]
    \centering
    \subfloat[
    \label{Mesh_01}]{%
      \includegraphics[width=0.9\textwidth]{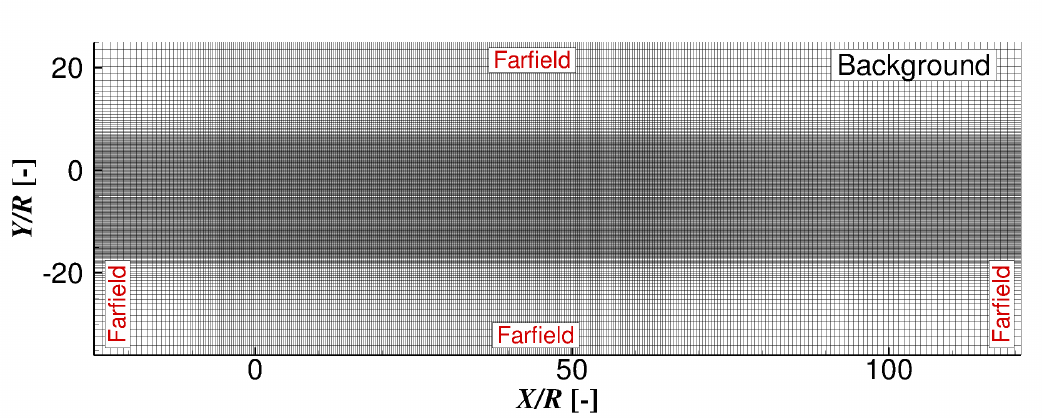}
    }\vspace{-4 mm}
    
    \subfloat[
    \label{Mesh_02}]{%
      \includegraphics[width=0.9\textwidth]{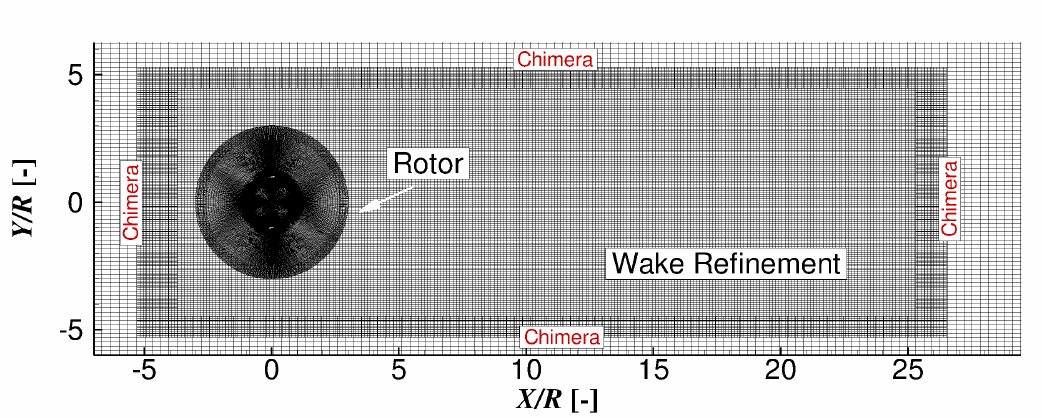}
    }\vspace{-4 mm}
    
    \subfloat[
    \label{Mesh_03}]{%
      \includegraphics[width=0.4\textwidth]{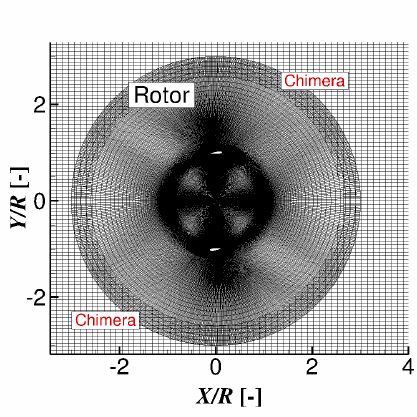}
    }   
    \subfloat[
    \label{Mesh_04}]{%
      \includegraphics[width=0.4\textwidth]{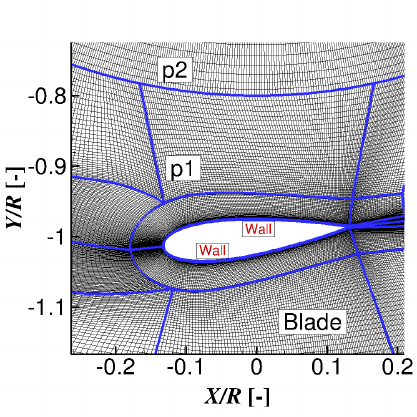}
    }    
    
    \subfloat[
    \label{Mesh_05}]{%
      \includegraphics[width=0.45\textwidth]{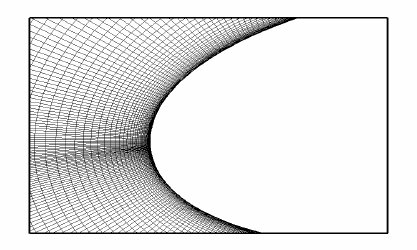}
    }  
    \subfloat[
    \label{Mesh_06}]{%
      \includegraphics[width=0.45\textwidth]{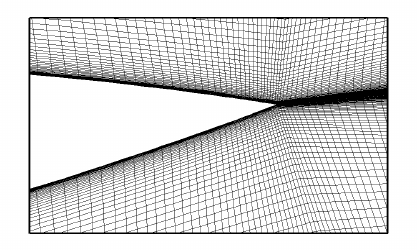}
    }       
    \caption{Computational mesh used in the simulations: background (\ref{Mesh_01}), wake refinement (\ref{Mesh_02}), rotor (\ref{Mesh_03}) and detailed mesh around the blade (\ref{Mesh_04}-\ref{Mesh_06}). The boundary conditions used are described by the red coloured text.}
   \label{Mesh}
  \end{figure}

The fully structured mesh technique was employed in the present studies. The mesh consists of three grid components, namely background, wake refinement and rotor meshes, as illustrated in Figure \ref{Mesh}. The grid overlapping (Chimera) technique was applied in the studies, enabling high quality meshes to be built separately for each grid component. This approach greatly simplifies the mesh generation. The background mesh is discretized by 329 x 193 grid points in $X$ and $Y$ directions, respectively. The domain size is -25$R$ x 120$R$ in $X$ direction and 50$R$ wide in $Y$ direction. A substantial mesh refinement was applied in the location where the rotor is located. Within this region, the wake refinement mesh consisting of an equi-distance cell size of $\Delta/R \approx$ 0.12 extending from $X/R$ = -5.3 up to 31.8 was introduced. This grid component consists of 257 x 141 grid points in $X$ and $Y$ directions, respectively. In the present studies, the resolution of this grid was increased up to 513 x 201 grid points ($\Delta/R \approx$ 0.06), and basic sensitivity studies will be carried out.

The rotor domain has a circular shape with the mesh component radius of 3$R$. The Chimera intersection area was defined near the outer radius for about 4 cells overlap. It is suggested not to apply the overlapping area too close to the blade wall to avoid unnecessary information loss from the data exchange between the fine mesh to the coarser meshes, especially when high velocity gradients need to be resolved such as flow involving massive separation. Figure \ref{Mesh_01} presents detailed mesh around the blade/airfoil. The number of grid points on the airfoil surface is 321. In wall normal direction, 65 grid points were applied up to point $p_1$ and 129 grid points up to point $p_2$, see Figure \ref{Mesh_01}. Among them, 32 cell layers are located within the boundary layer. These cell numbers are larger than the grid converged outcome of the study carried out by Song et al. \cite{song}. To properly resolve the boundary layer, the first grid point adjacent to the wall was set to meet the non-dimensional wall distance, $y^+$, less than unity. The averaged value of $y^+$ is around 0.15 which is sufficiently small.

To sum up, the mesh quality and quantity of the whole domains are described. Omitting the pole connector grid in the center of rotation of the rotor (origin), the maximum cell skewness is 0.75, which is small enough to avoid numerical errors due to high skewness cells influence. It shall be noted that the pole connector mesh was handled using the degenerate line boundary condition which allows a higher grid skewness on the pole boundary to be used. Several preceding works from the authors were carried out using this approach with good results \cite{bangga2, bangga3, bangga5, kim, bangga6}. The total number of grid points employed is of 327,524 divided into 92 different grid blocks. These grid blocks are further divided in the parallel computations of 24 CPUs carried out in the \textit{High Performance Computing Center Stuttgart} (HLRS).

The CFD simulations were carried out using a block-structured solver FLOWer from the German Aerospace Center (DLR) \cite{megaflow}. The code has been developed during the last years for wind turbine applications \cite{bangga2, bangga3, kim, bangga5}. The time integration is carried out by an explicit hybrid 5-stage Runge-Kutta scheme. Dual time-stepping according to Jameson \cite{jameson}, multigrid level 3 and implicit residual smoothing with variable coefficients were applied. The CFL numbers are 6.5 and 1.5 for the fine and coarse grid levels, respectively, in the multigrid scheme. The unsteady Reynolds-averaged Navier-Stokes (URANS) approach employing the Shear-Stress-Transport $k-\omega$ turbulence model according to Menter \cite{menter} was applied. The model has been widely used for industrial and academic purposes and is well known to be able to deliver good predictions for flows involving a strong adverse pressure gradient and separation. This has been also shown by previous works, for examples as in \cite{bangga1, bangga2, bangga3, bangga4, bangga5, bangga6, menter}. Three variants of time step sizes will be investigated, namely $\Delta t$ = $T/360$, $T/720$ and $T/1440$ which are equivalent to 1\degree{}, 0.5\degree{} and 0.25\degree{} blade rotation per physical time step, respectively. The solutions for each time step were iterated within 70 sub-iterations.

\FloatBarrier
\section{Results and Discussion} \label{Sect3}
In the present section, the results of the simulations for the two-bladed vertical axis wind turbine are presented and detailed discussions on the occurring physical phenomena are given. The discussions are divided into three main parts: Section \ref{Vali} presents the validation of the numerical results against available experimental data, and studies on temporal discretization sensitivity are presented. In Section \ref{Wake}, the wake characteristics downstream of the turbine are presented and discussed. At the end of the section, the dynamic stall phenomena at several tip speed ratios are evaluated in Section \ref{DS_S} showing the unsteady development of the flow during the rotor revolution.

\subsection{Validation and Temporal Discretization Studies} \label{Vali}
\begin{figure}[!t]
    \centering
    \subfloat{%
      \includegraphics[width=0.9\textwidth]{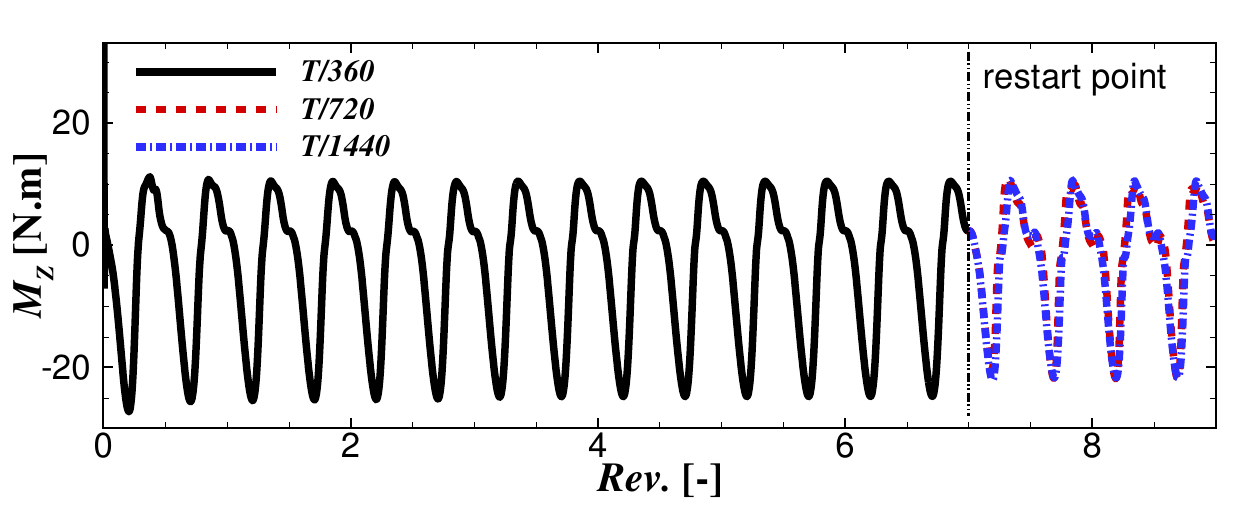}}
    \caption{Simulation strategy of the VAWT. The restart point was defined at 7th rotor revolution.}
   \label{Conv}
  \end{figure}

The simulations were carried out for 7 blade revolutions until the wake is fully developed at the time step size of $T/360$. Then the computations were restarted for 2 blade revolutions and smaller time step sizes ($T/720$ and $T/1440$) were applied as illustrated in Figure \ref{Conv}. This procedure was repeated for each simulated tip speed ratio. The power coefficient ($C_{power}$) was calculated based on the total moment coefficient around $Z$ axis ($M_Z$) defined as

\begin{equation}
C_{power} = -\frac{M_Z \Omega}{\frac{1}{2} \rho U_\infty^3 A}
\end{equation} 

\noindent where $\Omega$ is the rotor rotational speed in [rad/s], $\rho$ is the air density in [kg/m$^3$], $U_\infty$ is the wind speed of 8 m/s and $A$ is the cross section area of 2$R$. It shall be noted that positive power production is generated by negative driving moment for a rotor rotates in clockwise direction. In these studies, the results are time averaged only for the last rotor revolution.

\begin{figure}[!t]
    \centering
    \subfloat[
    \label{CP_01}]{%
      \includegraphics[width=0.45\textwidth]{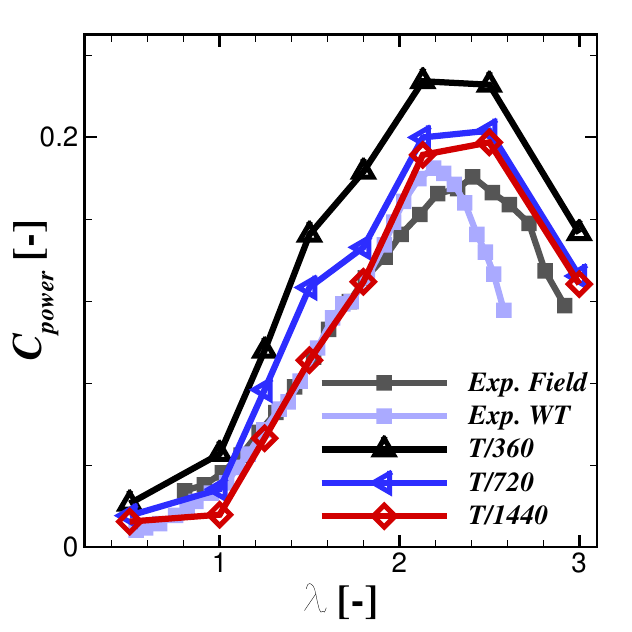}
    }
    \subfloat[
    \label{CP_02}]{%
      \includegraphics[width=0.45\textwidth]{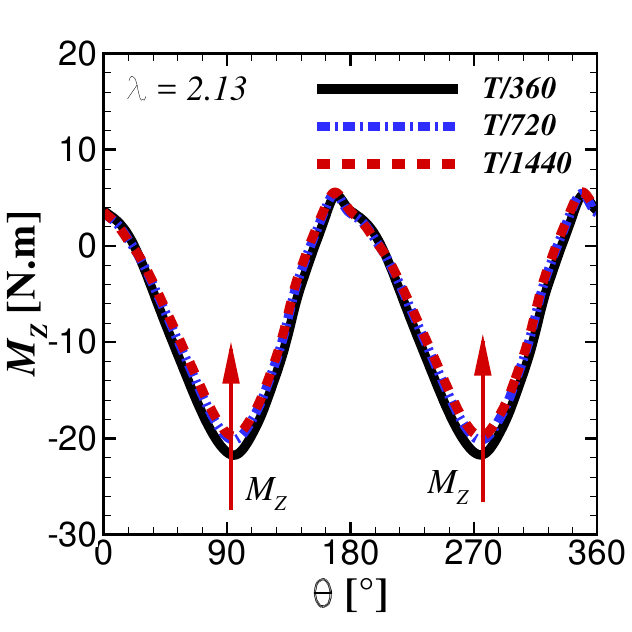}
    }  
    \caption{(\ref{CP_01})Comparison of the predicted power coefficient to measurement data. Finer time step size shows better agreement with experiment. (\ref{CP_02}) Moment around $Z$ axis for various time step sizes at $\lambda$ = 2.13. The computations were carried out applying the wake refinement grid for the resolution of 257 x 141 grid points.}
   \label{CP}
  \end{figure}

Figure \ref{CP_01} presents the comparison between the CFD simulations to the measurement data, both field (grey line) and wind tunnel tests (purple line), obtained from Li et al. \cite{li}. It can be seen that the wind tunnel and field test measurements exhibit very similar power curve at low tip speed ratios, approximately between $\lambda$ = 1.0 to 2.0, but a closer look on the figure shows that the power generation in the field measurement is a bit larger. A noticeable deviation between these two results can be observed for the tip speed ratio larger than the optimum power coefficient position ($\lambda \approx$ 2.19). It was documented in \cite{li} that the fluctuation amplitudes obtained by field test show larger value than the results of wind tunnel at low and high tip speed ratios. Li et al. \cite{li} argued that the deviation might occur as the result of the wind speed variation during the field test studies. It shall be noted that the same wind turbine was used in the wind tunnel and field measurements.
  
In Figure \ref{CP_01}, black line represents the CFD results employing the time step size of $T$/360, blue line of $T$/720 and red line of $T$/1440. The coarsest studied time scale results in an overestimation of the generated rotor power compared to measurement data. However, the general characteristics of the power curve is already reasonably captured. The prediction can be further improved by refining the time step size. As can be seen, the CFD predictions successively become more accurate as the time scale reduces to $T$/720 and $T$/1440. The computations even show excellent results for the time step size of $T$/1440 especially for the small tip speed ratio. The improvement of the $C_{power}$ prediction by refining the time step size is a result from the variation of $M_Z$ with $\Delta t$ depicted in Figure \ref{CP_02}. It can be seen that $M_Z$ becomes more positive and this causes reduction in the generated power production, alleviating the $C_{power}$ overestimation. Physically, the use of the smaller time step size increases the ability of the CFD computations in resolving the trailed vortices that are highly unsteady. This causes a stronger induction effect that further reduces the local angle of attack seen by the blade section, which in turn reduces the local power generation. Furthermore, the finer time steps seems to be able to capture unsteady separation especially for the small tip speed ratio. It is worth noting that, in general, unsteady simulations require a certain time step to resolve unsteady effects accurately. The wind tunnel experiment is, however, still over estimated by the CFD predictions for the larger tip speed ratio, but their magnitude is similar to the field measurement data. The discrepancy between the CFD results to the wind tunnel experiment is expected to stem from two main factors, (1) a smaller time step size is required and (2) three dimensional effects due to tip loss occur. The latter becomes reasonable since the studied blade has no taper, and it is already well known that finite aspect ratio reduced the blade aerodynamic performance which may result in the reduction of rotor power in the wind tunnel test.

\subsection{Wake Characteristics} \label{Wake}
\begin{figure}[!]
  \vspace{-5mm}
    \centering
    \subfloat[
    \label{Wake_01}]{%
      \includegraphics[width=0.175\textwidth]{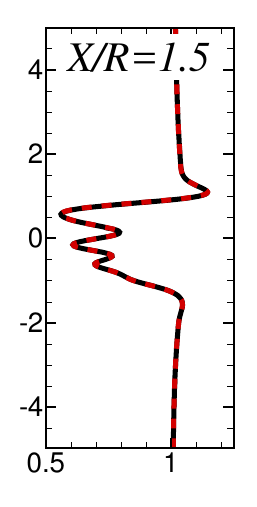}
    }
    \subfloat[
    \label{Wake_02}]{%
      \includegraphics[width=0.175\textwidth]{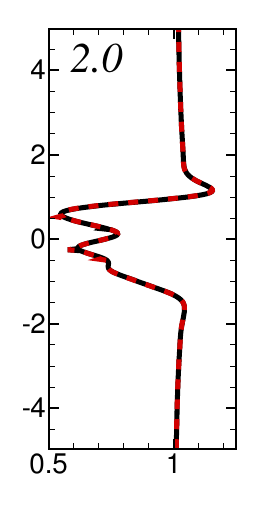}
    }
    \subfloat[  
    \label{Wake_03}]{%
      \includegraphics[width=0.175\textwidth]{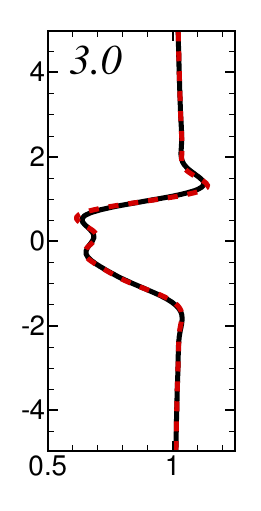}
    }
    \subfloat[
    \label{Wake_04}]{%
      \includegraphics[width=0.175\textwidth]{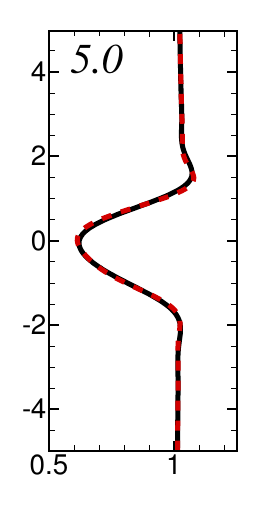}
    }    
    \subfloat[
    \label{Wake_05}]{%
      \includegraphics[width=0.175\textwidth]{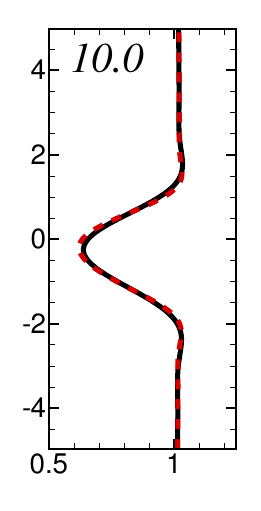}
    }        

    \vspace{-3mm}
    \subfloat[
    \label{Wake_06}]{%
      \includegraphics[width=0.175\textwidth]{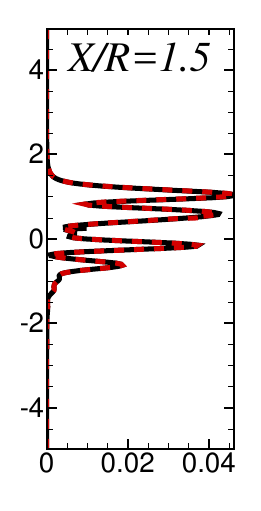}
    }
    \subfloat[
    \label{Wake_07}]{%
      \includegraphics[width=0.175\textwidth]{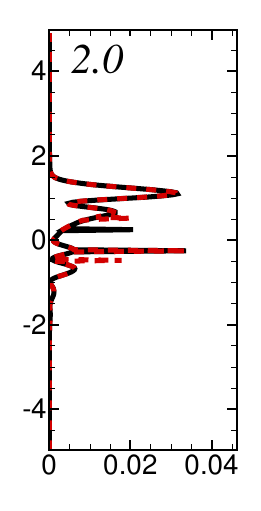}
    }  
    \subfloat[
    \label{Wake_08}]{%
      \includegraphics[width=0.175\textwidth]{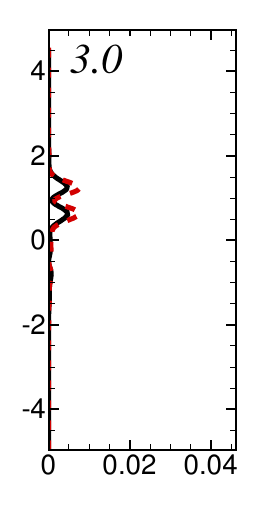}
    }
    \subfloat[
    \label{Wake_09}]{%
      \includegraphics[width=0.175\textwidth]{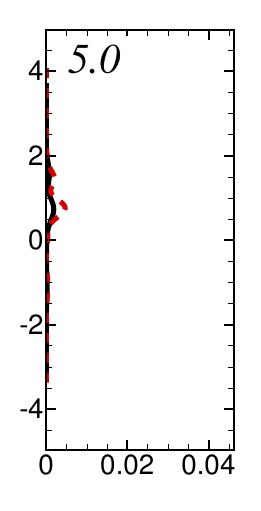}
    }    
    \subfloat[
    \label{Wake_10}]{%
      \includegraphics[width=0.175\textwidth]{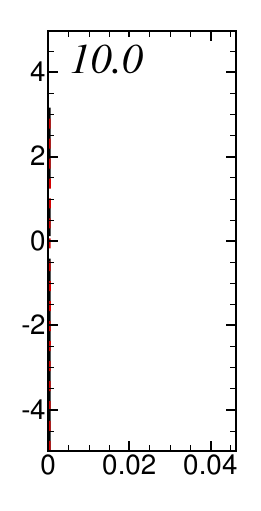}
    }
    
    \vspace{-3mm}
    \subfloat[
    \label{Wake_11}]{%
      \includegraphics[width=0.175\textwidth]{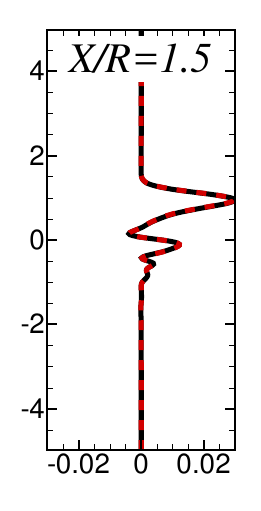}
    }
    \subfloat[
    \label{Wake_12}]{%
      \includegraphics[width=0.175\textwidth]{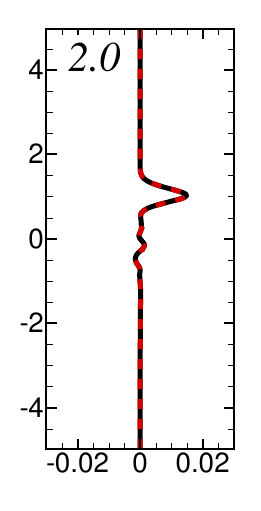}
    }  
    \subfloat[
    \label{Wake_13}]{%
      \includegraphics[width=0.175\textwidth]{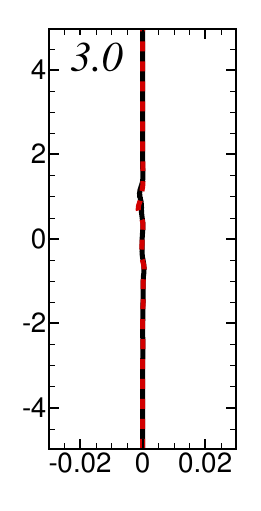}
    }
    \subfloat[
    \label{Wake_14}]{%
      \includegraphics[width=0.175\textwidth]{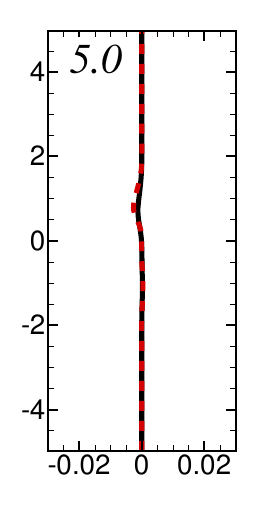}
    }    
    \subfloat[
    \label{Wake_15}]{%
      \includegraphics[width=0.175\textwidth]{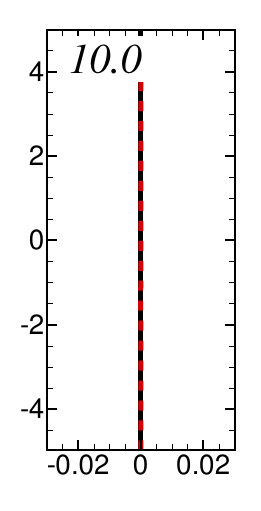}
    }            
    \caption{Unconditionally time averaged wake profiles for $\lambda$ = 1.50 at different downstream locations at $X/R$ = 1.5, 2.0, 3.0, 5.0 and 10.0 from successively left to right: mean streamwise velocity component $\overline{U}/U_{\infty}$ (top), mean streamwise variance $\overline{u'u'}/U_{\infty}^2$ (middle) and Reynolds shear stress component $\overline{u'v'}/U_{\infty}^2$ (bottom) evaluated from unsteady fluctuations in the wake. $y$ axis is the $Y$-direction in [m]. Solid-black line represents the CFD results for the wake refinement grid resolution of 257 x 141 grid points and dashed-red line is for 513 x 201 grid points. The employed time step is $T/1440$.}
   \label{Wake}
  \end{figure}

The mean velocity ($\overline{U}/U_{\infty}$), streamwise variance ($\overline{u'u'}/U_{\infty}^2$) and shear stress ($\overline{u'v'}/U_{\infty}^2$) profiles at several downstream locations are shown in Figure \ref{Wake}. As already briefly mentioned in Section \ref{Setup}, two different resolutions of the wake refinement mesh were employed, namely 257 x 141 and 513 x 201 grid points (streamwise x crosswise) for the baseline (black-solid line) and fine (red-dashed line) meshes, respectively. It is shown that both the grid resolutions produce very similar magnitude of the wake profiles, inferring that the mesh is spatially converged. Figures \ref{Wake_01}-\ref{Wake_05} show that the streamwise velocity profile changes according to its distance to the rotor plane. A strong wake deficit is observed close to the rotor and it gradually becomes fuller with increasing streamwise distance. Flow acceleration occurs near the tip area especially at $Y/R$ = 1.0. This seems to be related to the direction of the rotor rotation.  The wake velocity profile becomes nearly Gaussian at $X/R$ = 5.0, and this defines the near wake distance of the present case according to S{\o}rensen et al.\cite{sorensen}.

For the streamwise variance profile, strong peaks are observed for the whole rotor area at -1.0 $< Y/R <$ 1.0. This behaviour is in contrast to the wake characteristics for HAWTs where the peaks are strong only near the tip and hub areas \cite{kim}, inferring that the tip vortices are the main contributors to the turbulence mixing in the wake. In case of VAWTs, the interactions of wake vortices occur very early close to the rotor because of the blade vortex interaction phenomena (BVI) \cite{bangga1}. As no turbulence inflow is considered in the present studies, both $\overline{u'u'}/U_{\infty}^2$ and $\overline{u'v'}/U_{\infty}^2$ are caused mainly by periodic motion of the blade passage. It is shown that the Reynolds stresses magnitudes reduce with increasing streamwise distance. This confirms preceding studies that only the stochastic part of $\overline{u'v'}/U_{\infty}^2$ enhances the momentum mixing in the wake area \cite{kim, lignarolo}.

\subsection{Dynamic Stall Onset} \label{DS_S}
\begin{figure}[b]
    \centering
    \subfloat[Normal force
    \label{DS_01}]{%
      \includegraphics[width=0.45\textwidth]{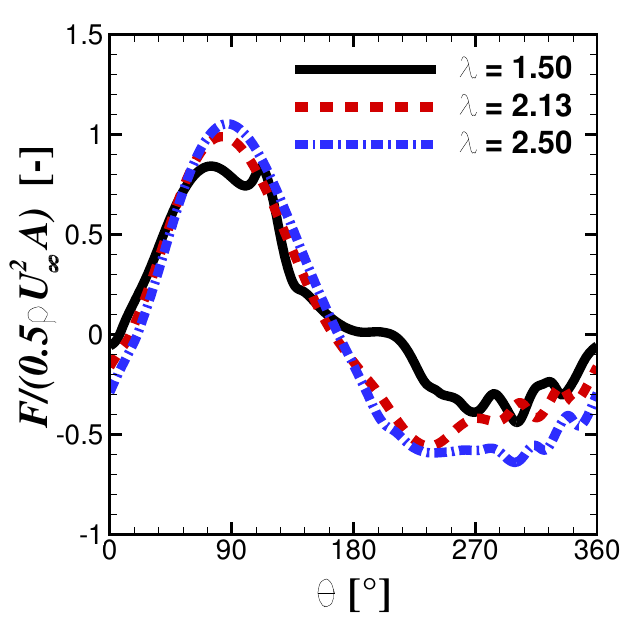}
    }
    \subfloat[Tangential force
    \label{DS_02}]{%
      \includegraphics[width=0.45\textwidth]{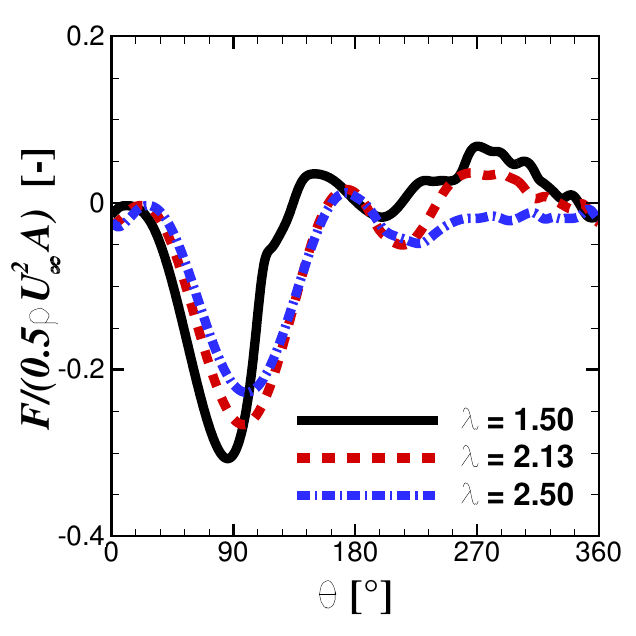}
    }  
    \caption{Normal ($F_N$) and tangential ($F_T$) forces acting on the blade (relative to chord) for several tip speed ratios.}
   \label{DS}
  \end{figure}

In this section, the effect of dynamic stall on the aerodynamic performance of VAWT will be discussed. Three different tip speed ratios: below ($\lambda$ = 1.50), near ($\lambda$ = 2.13) and above ($\lambda$ = 2.50) the optimum condition ($\lambda$ = 2.19) are evaluated. Figure \ref{DS} presents the normalized normal and tangential forces for these studied tip speed ratios. The forces direction is relative to the chord line. It shall be noted that the observation is made only for blade 1 employing the time step size of $T/1440$. 

It can be seen in Figure \ref{DS} that $F_N$ increases from $\theta$ = 0\degree{} until it reaches the maximum at $\theta$ = 90\degree{}. Then, it drops gradually and a strong non-linearity of the curve is observed for $\theta >$ = 200\degree{} indicating the stall region. These characteristics occur at a smaller azimuth angle for $F_T$. It seems that $F_N$ and $F_T$ obtained from both the higher tip speed ratios ($\lambda$ = 2.13 and 2.50) are very similar except in the stall area, but they differ strongly for the lower $\lambda$ case. This confirms that the dynamic stall influence becomes stronger with decreasing tip speed ratio \cite{bangga1, laneville, akins}.

\begin{figure}[!t]
    \centering
    \subfloat[$\theta$ = 0\degree{}
    \label{DS-FLOW_01}]{%
      \includegraphics[width=0.33\textwidth]{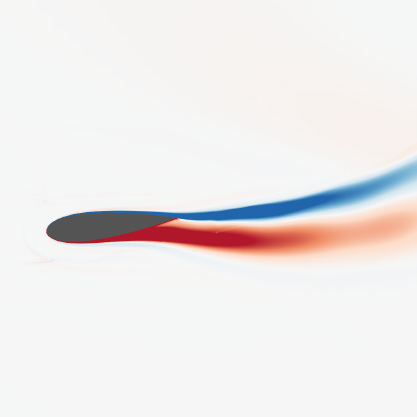}
    }
    \subfloat[$\theta$ = 72\degree{}
    \label{DS-FLOW_03}]{%
      \includegraphics[width=0.33\textwidth]{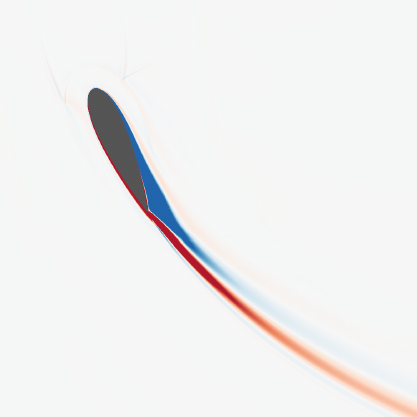}
    }
    \subfloat[$\theta$ = 100\degree{}
    \label{DS-FLOW_04}]{%
      \includegraphics[width=0.33\textwidth]{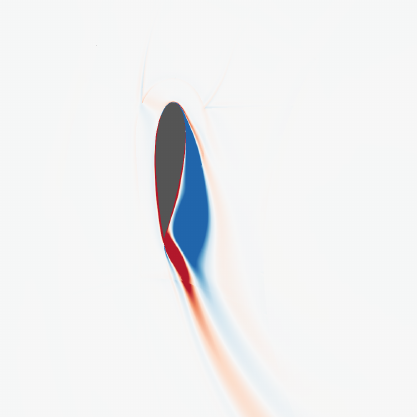}
    }
    
    \subfloat[$\theta$ = 112\degree{}
    \label{DS-FLOW_05}]{%
      \includegraphics[width=0.33\textwidth]{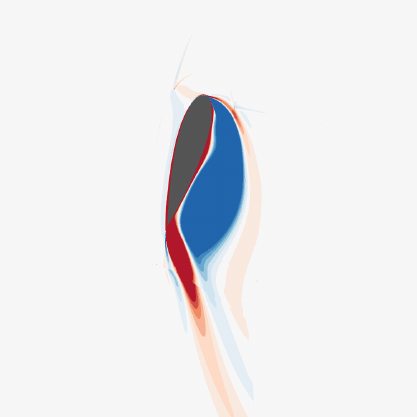}
    }
    \subfloat[$\theta$ = 132\degree{}
    \label{DS-FLOW_06}]{%
      \includegraphics[width=0.33\textwidth]{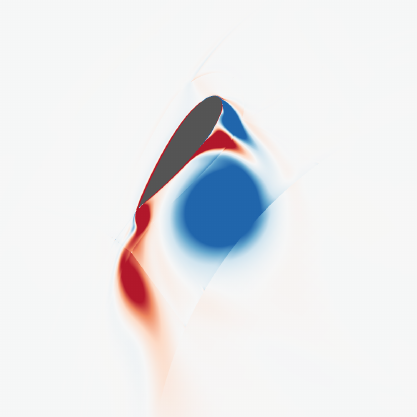}
    }  
    \subfloat[$\theta$ = 180\degree{}
    \label{DS-FLOW_07}]{%
      \includegraphics[width=0.33\textwidth]{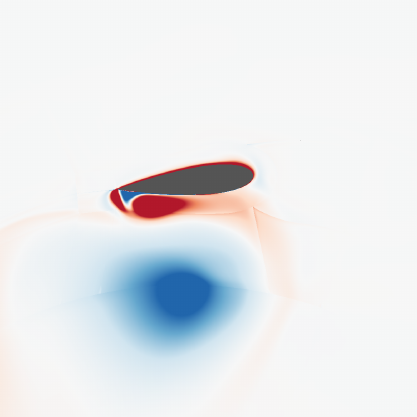}
    }
    
    \subfloat[$\theta$ = 206\degree{}
    \label{DS-FLOW_08}]{%
      \includegraphics[width=0.33\textwidth]{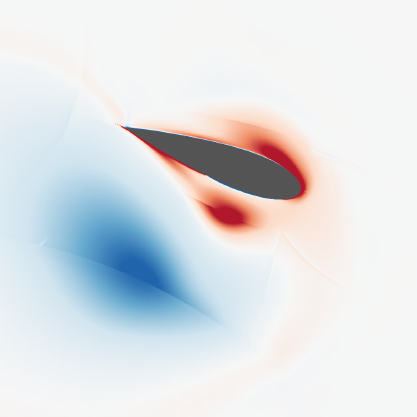}
    }
    \subfloat[$\theta$ = 266\degree{}
    \label{DS-FLOW_09}]{%
      \includegraphics[width=0.33\textwidth]{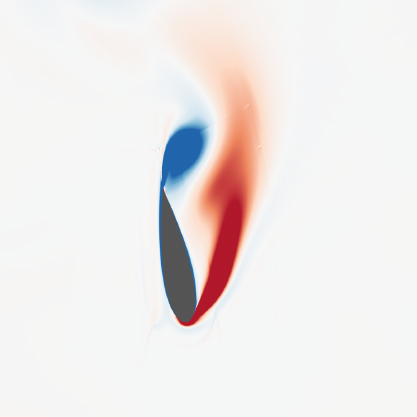}
    }
    \subfloat[$\theta$ = 332\degree{}
    \label{DS-FLOW_10}]{%
      \includegraphics[width=0.33\textwidth]{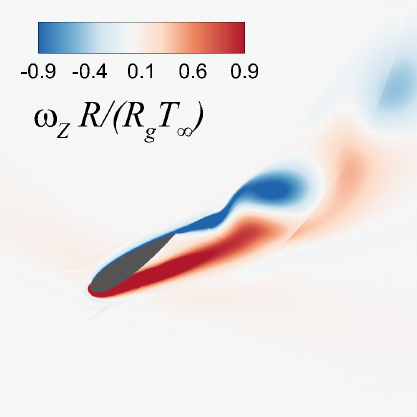}
    }    
    \caption{Crosswise vorticity field ($\omega_Z$) surrounding the blade during one rotor revolution for $\lambda$ = 1.50. It shall be noted that blue color indicates that the vortex direction is clockwise (CW) while red color is counter-clockwise (CCW). The magnitude is normalized by $R/(R_gT_\infty)$, where $R$, $R_g$ and $T_\infty$ are the blade radius (1 m), gas constant (287.1 J kg$^{−1}$K$^{−1}$) and free stream temperature (288.15 K), respectively.}
   \label{DS-FLOW}
  \end{figure}

Figure \ref{DS-FLOW} presents the dimensionless vorticity field around the blade for the tip speed ratio of 1.50, which was calculated from the $X$- and $Y$-velocity components as:

\begin{equation}
\omega_Z = \frac{\partial v}{\partial x} - \frac{\partial u}{\partial y}.
\end{equation}
 
\noindent Blue and red colors represent the shedding vortices in clockwise (CW) and counter-clockwise (CCW) directions, respectively. The observation is focused only on the flow development for blade 1 at various azimuth angles during one rotor revolution. The employed time step size is $T/1440$. For clarity, the blade section is divided into two parts: the inner side is the side of airfoil where the normal vector its surface points towards the center of rotation and the outer side is otherwise. It shall be noted that positive normal force is when the force acts towards the center of rotation, and positive tangential force is towards the trailing edge.

In Figure \ref{DS-FLOW_01}, at $\theta$ = 0\degree{}, two main vortices shedding from the blade surfaces are observed, which are skewed towards the outer blade surface. This phenomenon stems from the pitch angle effect causing non-symmetrical flows on the inner and outer blade sides to occur. As a consequence, negative normal and tangential force coefficients are observed. The effect seems to be stronger for the larger operating tip speed ratios. With increasing azimuth angle, now the vortices are skewed towards the opposite direction because the angle of attack seen by the blade section increases. This causes the normal force augmentation up to maximum at $\theta$ = 72\degree{} (for $\lambda$ = 1.50). Further increasing $\theta$ causes stronger trailing edge separation to occur on the inner side of the blade (Figure \ref{DS-FLOW_04}) indicated by a weak $F_N$ drops at $\theta$ = 100\degree{}. Slightly at a higher azimuth angle, at $\theta$ = 112\degree{}, displacement effect by the vortical structure on the inner side of the blade causes a local augmentation of $F_N$. Despite that, $F_T$ seems not too sensitive towards this effect where no local change at this azimuth position is observed in Figure \ref{DS_02}.

\begin{figure}[!t]
    \centering
    \subfloat[
    \label{I-FLOW_01}]{%
      \includegraphics[width=0.45\textwidth]{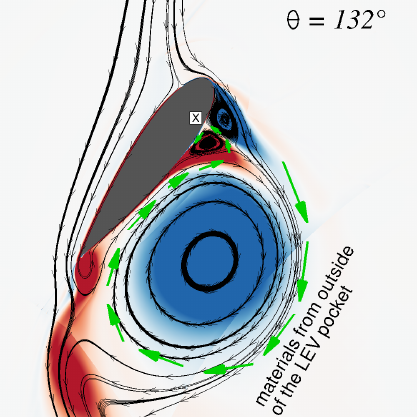}
    }
    \subfloat[
    \label{I-FLOW_02}]{%
      \includegraphics[width=0.45\textwidth]{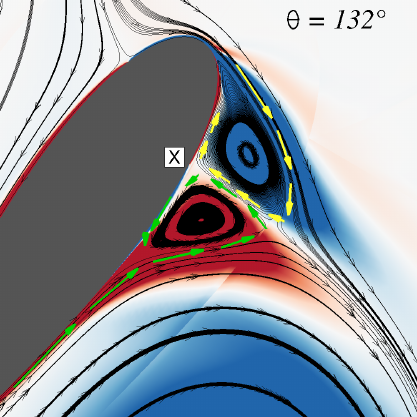}
    }
    
    \subfloat[
    \label{I-FLOW_03}]{%
      \includegraphics[width=0.75\textwidth]{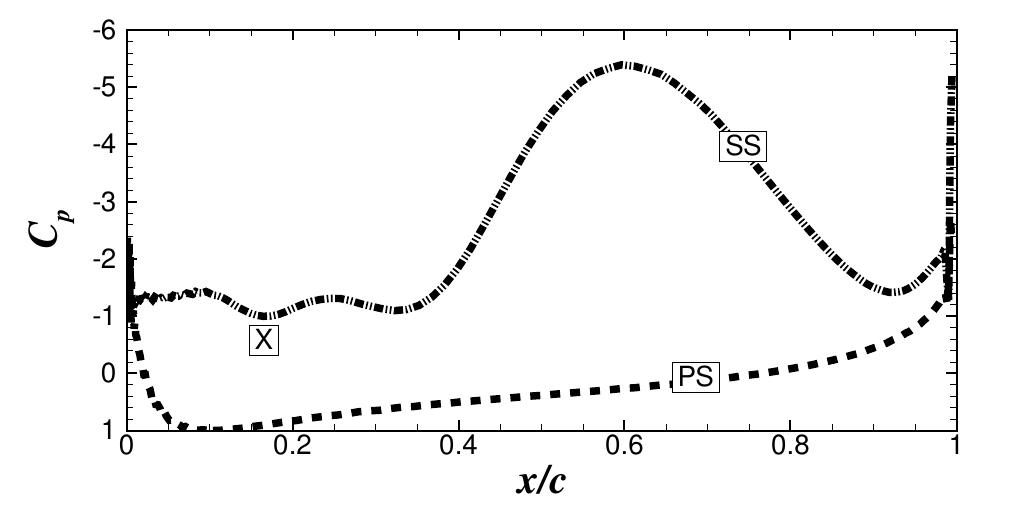}
    }  
    \caption{Flow insertion phenomenon colored by the crosswise vorticity (\ref{I-FLOW_01} and \ref{I-FLOW_02}). The same color level as in Figure \ref{DS-FLOW} is used. Figure \ref{I-FLOW_02} shows the resulting pressure distribution. PS and SS indicate pressure and suction sides, respectively.}
   \label{I-FLOW}
  \end{figure}

A sharp drop of the blade forces occurs for the higher azimuth angle up to $\theta$ = 132\degree{}. The origin is quite obvious as the clockwise vortical structure start to detach from the blade surface, depicted in Figure \ref{DS-FLOW_05}. It has been documented by many authors that deep dynamic stall happens when the breakdown of the travelling vortex takes place \cite{hutomo, bangga1, bangga4, mccroskey, leishman}. The clockwise vortex breakdown (blue color) is caused by the flow insertion effect by the counter-clockwise vortex (red color) near the leading edge. This supports the numerical studies carried out for the other turbines in \cite{hutomo, bangga1}. For clarity of the discussion, Figure \ref{I-FLOW} presents the enlarged view of Figure \ref{DS-FLOW_05}, and flow streamlines surrounding the blade airfoil are plotted. The phenomenon starts when flow materials outside of the leading edge vortex (LEV) pocket are transported towards the leading edge in the near wall area creating a small but intense counter-clockwise vortex a bit downtsream of the leading edge, see green arrows in Figure \ref{I-FLOW}. At the same time, a secondary clockwise vortex is formed near the leading edge, see yellow arrows in Figure \ref{I-FLOW}. This causes the main LEV pocket to separate from the airfoil body, causing the sudden drop of the blade forces. In addition, the flow direction of the observed counter-rotating secondary LEVs is almost normal to the wall at point X (see Figure \ref{I-FLOW_02}). This results in the increased pressure effect, as depicted in Figure \ref{I-FLOW_03}, that has a significant contribution in the aerodynamic losses of the blade.

The normal force gradually decreases up to $\theta$ = 270\degree{}. On the other hand, the tangential force becomes more positive within these azimuth angles except at $\theta$ = 180\degree{} where a local drop is observed, which seems to be connected to the generation of the remarkable counter-clockwise vortex near the trailing edge in Figure \ref{DS-FLOW_06}. For the azimuth angles higher than 270\degree{}, the flow field is dominated by shedding vortices from the inner and outer sides of the blade that cause loads fluctuation to occur.

Generally, it can be seen that positive power generation (negative $F_T$) for blade-1 occurs within the azimuth angles of 0\degree{} $< \theta <$ 128\degree{} for $\lambda$ = 1.50, 0\degree{} $< \theta <$ 234\degree{} for $\lambda$ = 2.13 and almost the whole azimuth angle range (0\degree{} $< \theta <$ 360\degree{}) for $\lambda$ = 2.50. This indicates that the aerodynamic benefit is obtained for the larger operating tip speed ratio, even for $\lambda$ greater than the optimum value. Furthermore, setting the operating tip speed ratio to a larger one may increase the life cycle of the turbine because the dynamic stall and load fluctuation effects are alleviated.

\begin{figure}[!t]
    \centering
    \subfloat[
    \label{CPX_01}]{%
      \includegraphics[width=0.45\textwidth]{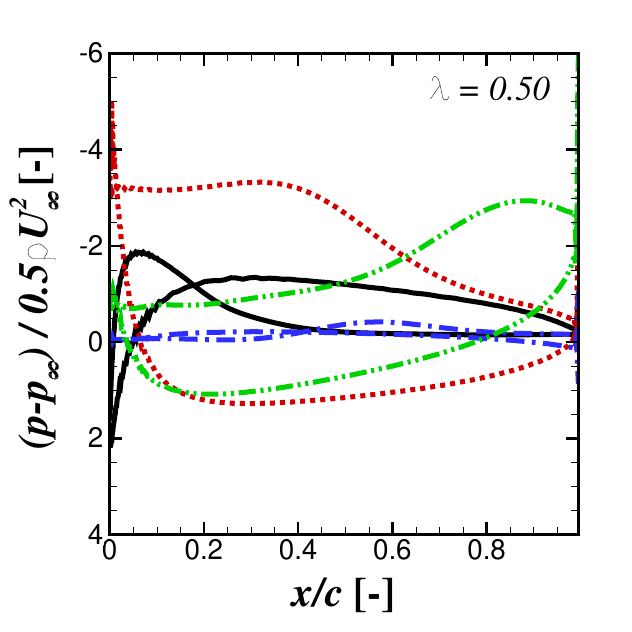}
    }
    \subfloat[
    \label{CPX_02}]{%
      \includegraphics[width=0.45\textwidth]{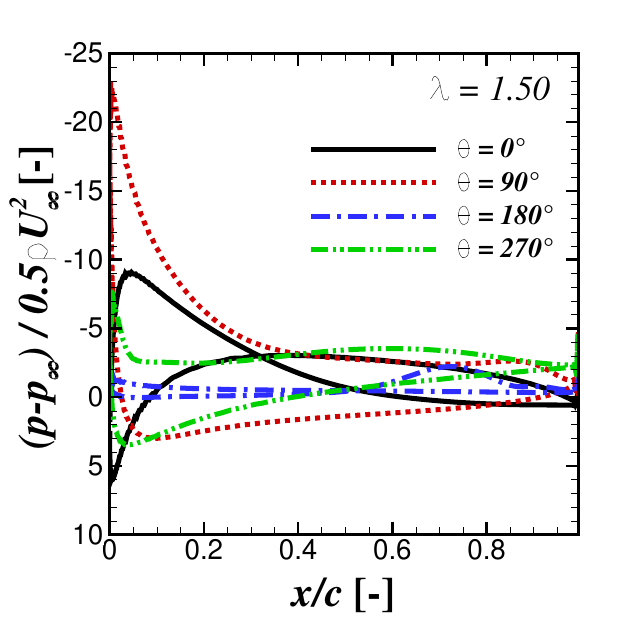}
    }  
    
    \subfloat[
    \label{CPX_03}]{%
      \includegraphics[width=0.45\textwidth]{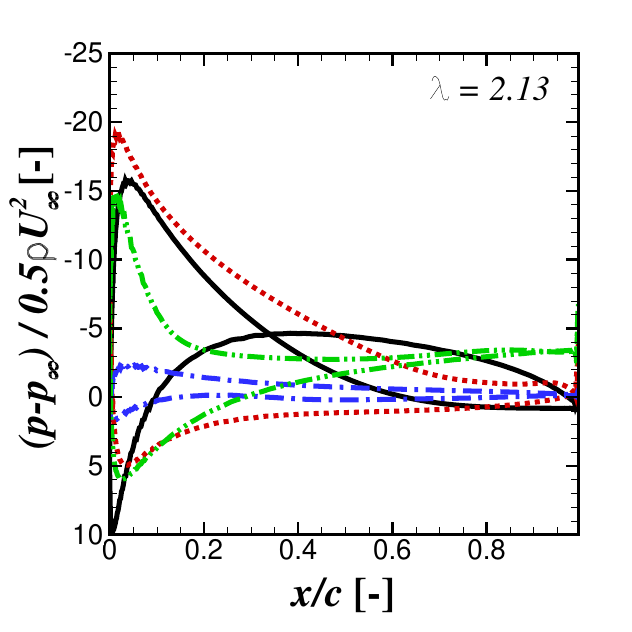}
    }
    \subfloat[
    \label{CPX_04}]{%
      \includegraphics[width=0.45\textwidth]{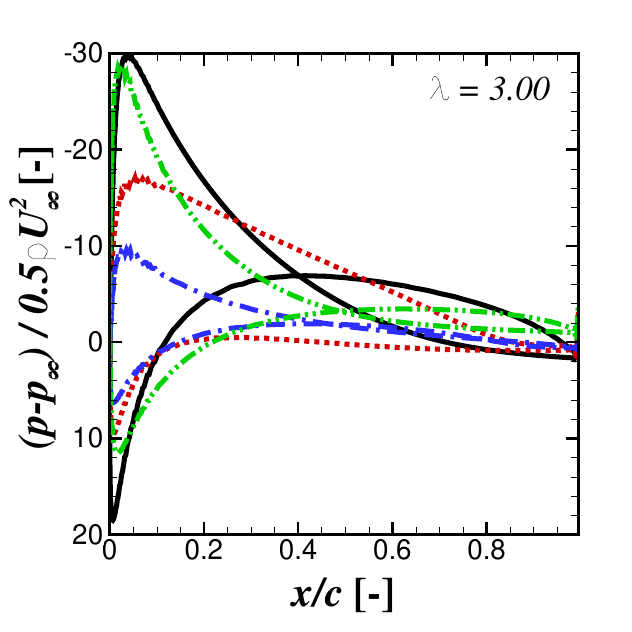}
    }      
    \caption{Dimensionless pressure acting on the blade surface normalized by the free-stream variables in the inertial frame of reference for several $\lambda$ and $\theta$. The rotational speed is excluded in the normalization to clearly show the magnitude of the pressure distribution.}
   \label{CPX}
  \end{figure}

In Figure \ref{CPX}, the distributions of pressure are presented for four different tip speed ratios and azimuth angles. The above discussions are further clarified as the shape of the pressure distributions seems to become more similar to the static airfoil conditions with increasing tip speed ratio. Furthermore, flow separation and variations of the pressure distribution shape between different azimuth angles are also alleviated. This is natural because the range of operating angle of attack reduces with $\lambda$, which in turn decreases the possibility of the blade to operate in the stall conditions.


\FloatBarrier
\section{Conclusion and Outlook} \label{section4}
Numerical simulations using 2D computational fluid dynamics (CFD) approaches have been
carried out in the present studies on a two-bladed vertical axis wind turbine (VAWT) for various tip speed ratios (0.50$< \lambda <$ 3.0). The radius of the rotor is 1 m with the chord length of 0.265 m, and is operating at a constant wind speed of 8 m/s. The numerical
results were compared to the experimental data by Li et al. \cite{li} to assess the accuracy of the computations. Three time step sizes of $T/360$, $T/720$ and $T/1440$ were evaluated, where $T$ represents the period of the rotor revolution. It was shown that the numerical results become more accurate for the smaller time step size. An excellent agreement with the experiment is obtained for the small tip speed ratio up to the optimum power coefficient at $\lambda$ = 2.19 even though flow separation occurs. For the larger tip speed ratios, the numerical results overestimate the wind tunnel experiment, but they are very similar to the field measurement data. The discrepancy is expected to arise due to tip loss effects that are not captured in the 2D simulations and a smaller time step size may be required to model the small vortices. 

The characteristics of wake are presented in the present studies in terms of the steamwise velocity ($U$), streamwise variance ($u'u'$) and Reynolds shear stress ($u'v'$) profiles. Two wake refinement grid resolutions were investigated, showing that the grid is spatially converged. The wake profile becomes Gaussian at $X/R$ = 5 indicating the location of the near wake distance for the studied case. It was shown that a remarkable flow acceleration occurs near the tip area especially at $Y/R$ = 1.0 that is connected to the direction of the rotor revolution. Strong peaks of the Reynold stress components ($u'u'$ and $u'v'$) are observed within the rotor area at -1.0 $< Y/R <$ 1.0 because noticeable momentum mixing occurs as a result from the blade vortex interaction. Additionally, it was shown that the strength of the Reynold stress components reduces with streamwise distance.

The flow field surrounding the blade during one rotor revolution and its impact on blade loads were evaluated. Complex flow field phenomena were observed especially at the azimuth angle larger than 90\degree{}, following the dynamic stall onset caused by the flow insertion effect. Power production for of the blade depends on the operating tip speed ratio. The azimuth range of the positive generated power from blade 1 increases with increasing $\lambda$. This indicates that designing VAWTs at higher tip speed ratios may be beneficial in term of power production, and this may can improve the life-time of the rotor since the dynamic stall effect is reduced.

For the future works, several aspects can be suggested in the field of VAWT studies. Fully resolved 3D simulations may improve the accuracy of the results especially in capturing the tip loss effects at high tip speed ratios that are not captured in the present works. Furthermore, higher fidelity simulation approaches like Large Eddy Simulation (LES) or Detached-Eddy Simulation (DES) are recommended for this purpose. Additionally, the performance assessment of a multi-rotor system is suggested in the subsequent works.

\section*{acknowledgements}
The authors gratefully acknowledge these following institutions for the supports: Ministry of Research, Technology and Higher Education of Indonesia for the funding through Directorate General of Higher Education (DGHE) scholarship and the High Performance Computing Center Stuttgart (HLRS) for providing computational time in the CFD simulations. This paper is dedicated to F.F.S. Lesy who passed away on June 12, 2017 while accompanying the author G. Bangga in finishing the works.

\FloatBarrier
\bibliographystyle{unsrt}
\bibliography{bangga_etal}

\end{document}